\documentclass[runningheads]{llncs}

\usepackage{booktabs}
\usepackage{enumitem}
\usepackage{graphicx}
\usepackage{multirow}
\usepackage{textcomp}
\usepackage{xspace}
\usepackage{hyperref}
\usepackage{url}
\usepackage{makecell}

\usepackage{pgfplots}
\usepgfplotslibrary{colorbrewer}
\pgfplotsset{compat=1.17}

\usepackage{numprint}

\usepackage[acronym, nowarn]{glossaries}
\makeglossaries
\newacronym{RFS}{RFS}{Recurrence-Free Survival}
\newacronym{RMSE}{RMSE}{root mean squared error}

\begin{document}

\title{MLC at HECKTOR 2022: The Effect and Importance of Training Data when Analyzing Cases of Head and Neck Tumors using Machine Learning}

\newcommand{\simulamet}{SimulaMet, Norway}
\newcommand{\uit}{UiT The Arctic University of Norway, Norway}
\newcommand{\oslomet}{Oslo Metropolitan University, Norway}

\newcommand{\decimalplaces}{3}

\author{%
    Vajira Thambawita\inst{1} \and 
    Andrea M. Storås\inst{1,2} \and 
    Steven A. Hicks\inst{1} \and \\ 
    Pål Halvorsen\inst{1,2} \and
    Michael A. Riegler\inst{1,3}
}

\authorrunning{Thambawita et al.}
    \institute{%
    \simulamet 
    \and
    \oslomet 
    \and
    \uit 
}

\maketitle

\begin{abstract}
Head and neck cancers are the fifth most common cancer worldwide, and recently, analysis of Positron Emission Tomography (PET) and Computed Tomography (CT) images has been proposed to identify patients with a prognosis. Even though the results look promising, more research is needed to further validate and improve the results. This paper presents the work done by team MLC for the 2022 version of the HECKTOR grand challenge held at MICCAI 2022.
For Task 1, the automatic segmentation task, our approach was, in contrast to earlier solutions using 3D segmentation, to keep it as simple as possible using a 2D model, analyzing every slice as a standalone image. In addition, we were interested in understanding how different modalities influence the results. We proposed two approaches; one using only the CT scans to make predictions and another using a combination of the CT and PET scans.
For Task 2, the prediction of recurrence-free survival, we first proposed two approaches, one where we only use patient data and one where we combined the patient data with segmentations from the image model. For the prediction of the first two approaches, we used Random Forest. In our third approach, we combined patient data and image data using XGBoost. Low kidney function might worsen cancer prognosis. In this approach, we therefore estimated the kidney function of the patients and included it as a feature. 
Overall, we conclude that our simple methods were not able to compete with the highest-ranking submissions, but we still obtained reasonably good scores. We also got interesting insights into how the combination of different modalities can influence the segmentation and predictions.
\end{abstract}

\section{Introduction}\label{section:introduction}
\begin{figure}[!t]

    \newcommand{\imagesize}{.16\linewidth}
    \centering
    \includegraphics[width=\imagesize, height=\imagesize]{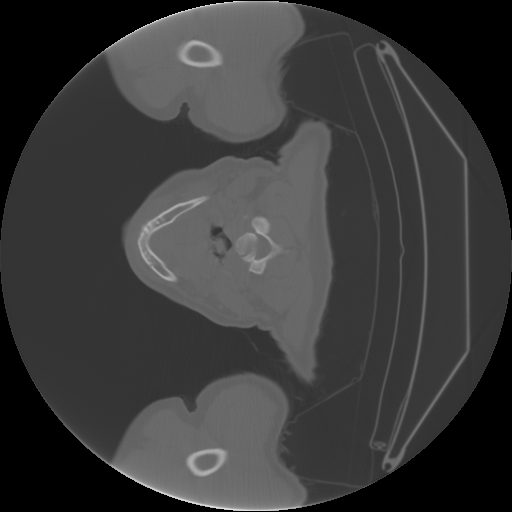}
    \includegraphics[width=\imagesize, height=\imagesize]{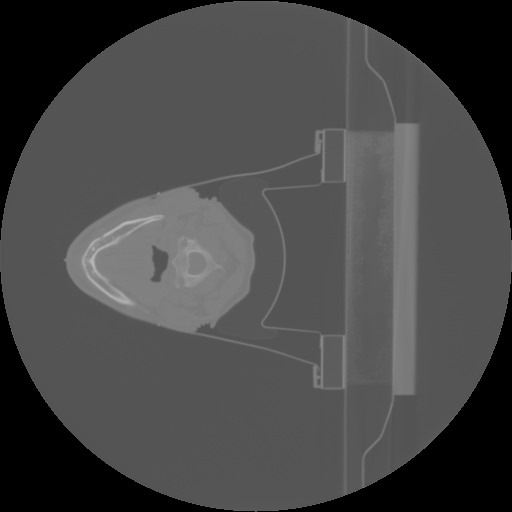}
    \includegraphics[width=\imagesize, height=\imagesize]{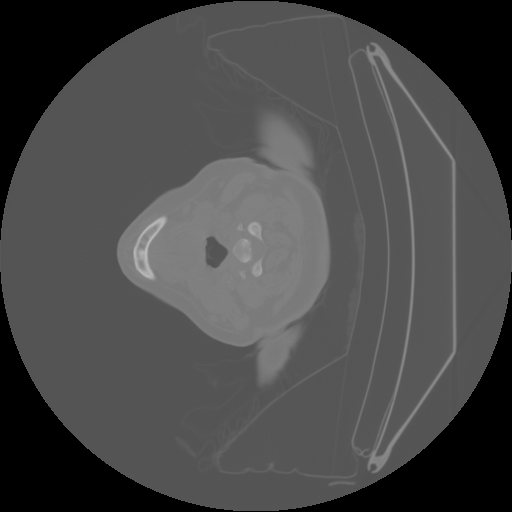}
    \includegraphics[width=\imagesize, height=\imagesize]{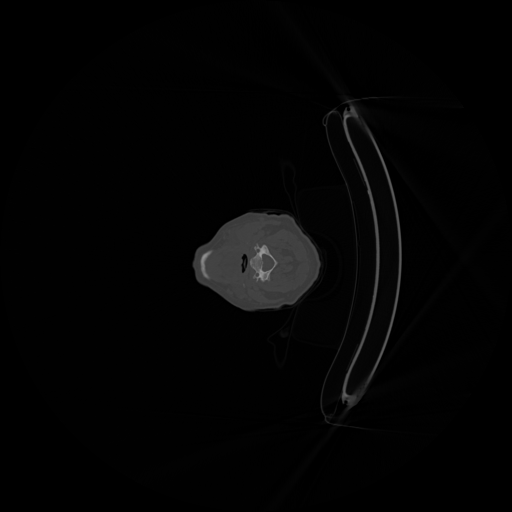}
    \includegraphics[width=\imagesize, height=\imagesize]{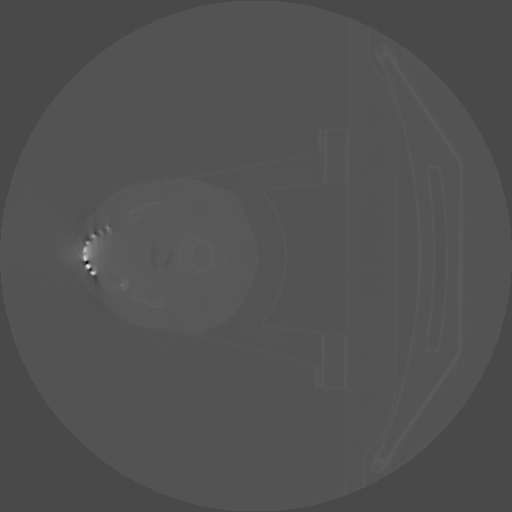}
    \includegraphics[width=\imagesize, height=\imagesize]{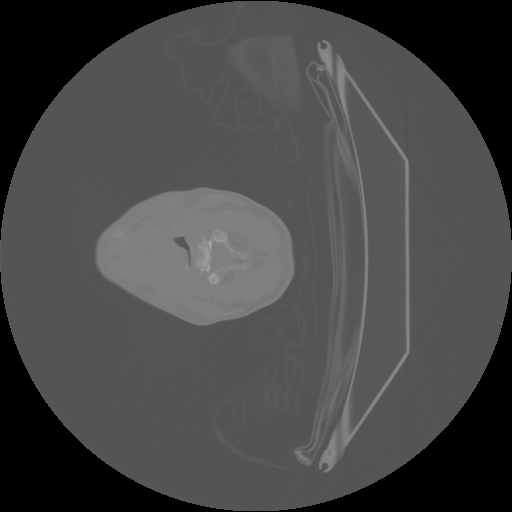}
    
    \includegraphics[width=\imagesize, height=\imagesize]{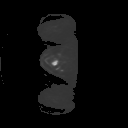}
    \includegraphics[width=\imagesize, height=\imagesize]{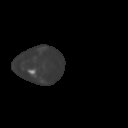}
    \includegraphics[width=\imagesize, height=\imagesize]{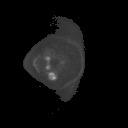}
    \includegraphics[width=\imagesize, height=\imagesize]{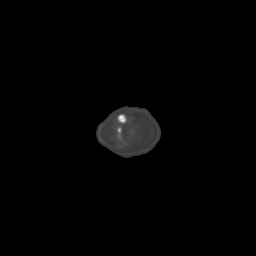}
    \includegraphics[width=\imagesize, height=\imagesize]{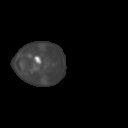}
    \includegraphics[width=\imagesize, height=\imagesize]{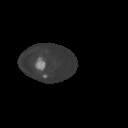}
    
    \includegraphics[width=\imagesize, height=\imagesize]{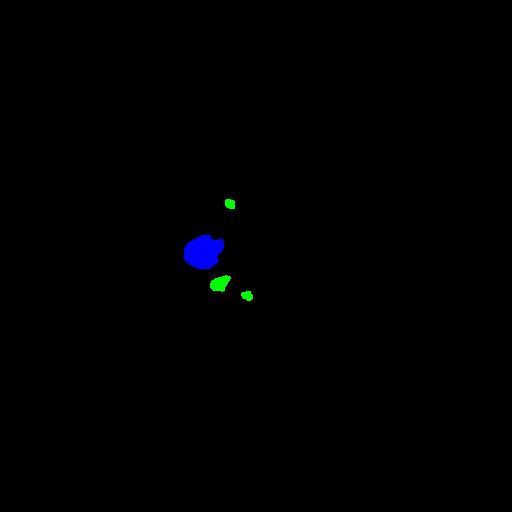}
    \includegraphics[width=\imagesize, height=\imagesize]{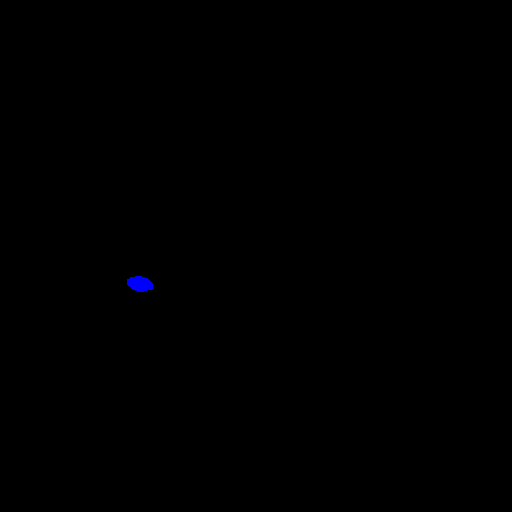}
    \includegraphics[width=\imagesize, height=\imagesize]{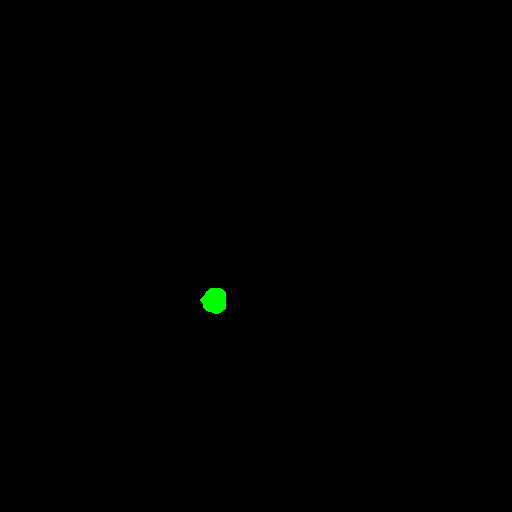}
    \includegraphics[width=\imagesize, height=\imagesize]{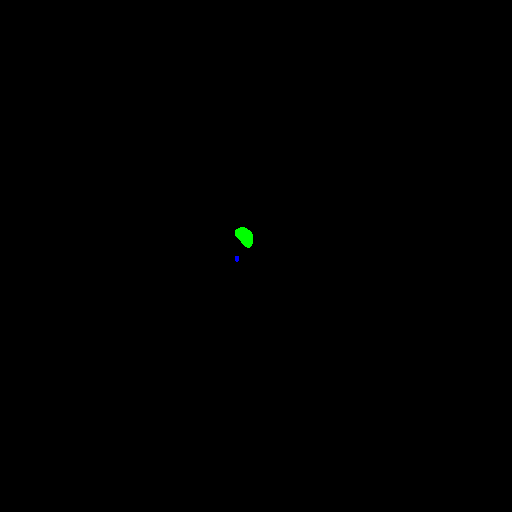}
    \includegraphics[width=\imagesize, height=\imagesize]{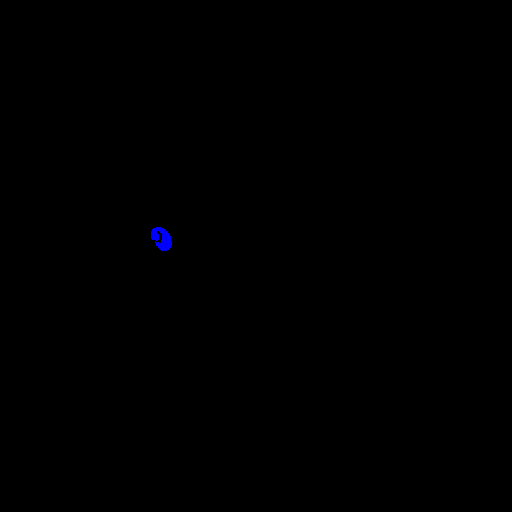}
    \includegraphics[width=\imagesize, height=\imagesize]{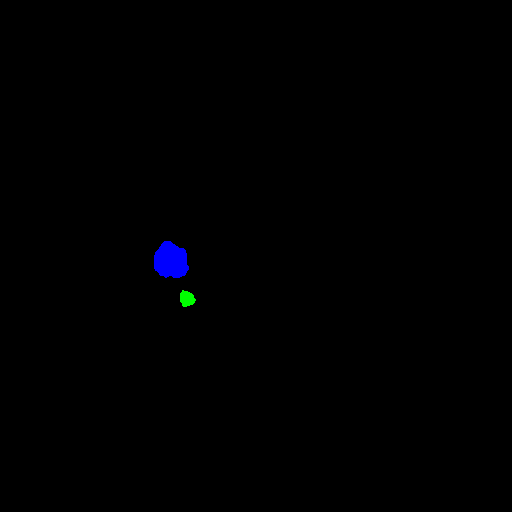}
    
    \caption{Six example data points from the development dataset provided by the HECKTOR organizers. The first row contains a slice from one of the CT images, the second row contains a slice from one of the PET images, and the third row contains the corresponding segmentation mask. Please note that the brightness of the PET images has been adjusted so that the contents are more easily visible.}
    \label{figure:Dataset_Examples}
\end{figure}

Head and neck cancers are among the most common cancer types worldwide. Early detection is critical as the tumor's size on diagnosis will dictate the patient's quality of life and chances of survival~\cite{Gerstner:2008}. Medical image analysis and radiomics have shown promising results in detecting different diseases and cancers~\cite{mittmann2022deep,duran2022prostattention,bandyk2021mri}, including those found in the head and neck~\cite{ren2021comparing,wahid2022evaluation,outeiral2022strategies}. In this paper, we describe our approaches for the HEad and neCK TumOR (HECKTOR) grand challenge held at MICCAI 2022~\cite{andrearczyk2020HecktorOverview,oreiller2022PaperHecktor}. Of the two tasks presented at the challenge, we participated in both. In Task 1, the aim was to segment tumors from Computed Tomography (CT) and Positron Emission Tomography (PET) scans of the head and neck (examples shown in Figure~\ref{figure:Dataset_Examples}). Task 2 asked for the prediction of \gls{RFS} based on clinical information about the patients presented in a tabular format, which also could be combined with the outputs from Task 1.

As the provided dataset contained different types of data, our strategy to tackle the HECKTOR challenge was to explore how the inclusion and combination of different modalities change the prediction outcome. In this respect, we investigated how CT and PET scans can be used individually or combined for tumor segmentation in Task 1 and how \gls{RFS} can be predicted using the meta-data with or without tumor information for Task 2. The main contributions of this paper are as follows:
\begin{enumerate}
    \item A comparison of simple segmentation methods using CT or PET slices individually or combined.
    \item Understanding the effect of combining different data modalities on the analysis results.
    \item Analysis of what features were most relevant for predicting \gls{RFS} using patient-related data and image features.
\end{enumerate}

\section{Methods}

In this section, we describe the methods we applied to solve Task 1 and 2, respectively. 

\subsection{Task 1: Segmentation of CT and PET scans}
\begin{figure}[!t]
    \centering
    \includegraphics[scale=0.7]{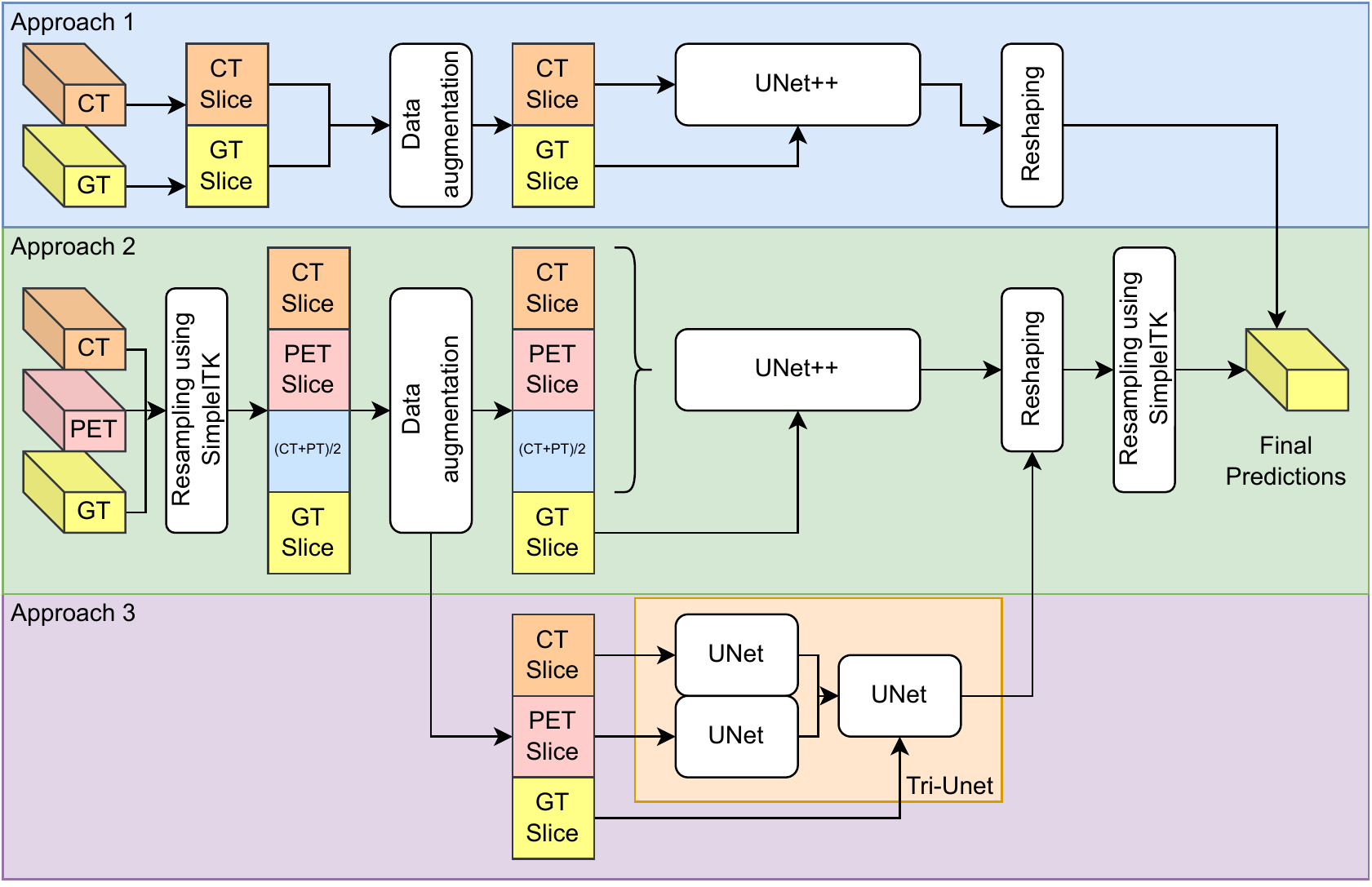}
    \caption{Three approaches used for Task 1. Approach 1: uses only CT images and the corresponding ground truth (GT). Approach 2: input stack of CT, PET, mean of CT and PET. Approach 3: use two separate UNet models for CT and PET and another UNet for final predictions. Reshaping sizes used in Approach 1 is different from the sizes used for Approach 2 and 3.}
    \label{fig:task_1_approaches}
\end{figure}

For Task 1, we used the provided development dataset consisting of CT scans, PET scans, and corresponding segmentation masks. we experimented with three different approaches as follows (see Figure~\ref{fig:task_1_approaches}):

\begin{description}
    \item[Approach 1:] Only using individual slices of the CT scans to predict tumors with a UNet++-based model~\cite{zhou2019unet++}.
    \item[Approach 2:] Combining the CT and PET scans by stacking CT, PET, and the mean of CT and PET images channel-wise and passing them through a UNet++-based model.
    \item[Approach 3:] Analyzing CT and PET slices separately in an ensemble-like setup using a TriUnet-based model~\cite{Thambawita++2021}.
\end{description}

These three approaches utilize the data provided in the HECKTOR competition differently, from simple to more complex. The following sections describe all the steps of data pre-processing, sampling, augmentation, implementation of the models, and post-processing.  

\subsubsection{Image Data pre-processing:}
We divided the development dataset into a training and a validation dataset containing $90\%$ and $10\%$ samples, respectively. For Approach 1, we extracted the slices from the CT and ground truth as \textit{.png} images without applying re-sampling because the shape of the CT and provided ground truth were the same. However, for Approaches 2 and 3, we performed slice extraction after re-sampling (using SimpleITK~\cite{yaniv2018simpleitk}). We used the same re-sampling as provided by the task organizers\footnote{\url{https://github.com/voreille/hecktor/blob/master/src/resampling/resample_2022.py}} with default spacing $(2,2,2)$. In addition, we normalized all CT and PET images into the range between $0$ and $255$, but not ground truth images that contain pixel values of $[0,1,2]$. After the extraction process, we noticed that the training dataset contained large number of true negative samples. Therefore, to avoid bias, we re-balanced the training dataset by extracting only slices with true positive pixels for H\&N Primary tumors (GTVp) and H\&N nodal Gross Tumor (GTVn). The class rebalancing was done by combining an equal number of true positive slices with the true negative slices extracted from the initial training dataset. To make a challenging validation dataset, we extracted only slices with GTVp and GTVn from the validation images. 

We applied similar image augmentation for all three approaches. The Albumentations~\cite{info11020125} library provides a set of augmentation options for image segmentation tasks. More information about the input parameters of the augmentation methods can be found in our GitHub repository\footnote{\url{https://github.com/vlbthambawita/hecktor\_2022\_MLC}}.    

\subsubsection{Model architectures, hyperparameters and inputs:}
The models for Task 1 were implemented in Pytorch~\cite{NEURIPS2019_9015} using the Segmentation Models library~\cite{Iakubovskii:2019}. All models were trained for $100$ epochs on hardware consisting of two Nvidia RTX 3080 Graphic Processing Units (GPUs) with 10 GB of memory each, an AMD Ryzen 9 3950X 16-core processor, and 64 GB memory. Submissions were made with the best-performing checkpoints, which were selected based on the performance on the validation dataset. For the first $50$ epochs, the learning rate was set to $0.0001$, then reduced to $0.00001$ for the remaining $50$ epochs. The Adam optimizer~\cite{kingma2014adam} with default parameters except the learning rate was used for all the experiments. Furthermore, we have used DiceLoss with skipped channel $0$ as the main loss function in the training process and the Intersection over Union (IoU) as a metric to evaluate our models. 

In Approach 1, we have used a UNet++-based model with $se\_resnext50\_32x4d$ as the encoder. The model was trained using only single channel CT input images and the corresponding ground truth masks after resizing them into $256\times256$ in the augmentation step. 

For Approach 2, we re-sampled the CT and PET slices and trained a UNet++ model. For this approach, we stack a CT slice, a PET slice, and the mean of the CT and PET slice in the color channel and use these as input to the model. The main objective of the second approach is to gain more information about using a UNet++-based architecture without making any major changes from the first approach. 

Approach 3 used a different architecture, TriUnet, which we introduced in our previous study~\cite{Thambawita++2021}. In this model, we input re-sampled single channel CT slices into one UNet~\cite{ronneberger2015u} and PET slices into another UNet. Then, the output of the two networks was passed through another UNet model, which accepts six input channels (3 channels output from the first and second UNet model for representing three classes of the ground truth). We used the same hyperparameters and trained the network as a single model using a single back-propagation step. The reason for not using UNet++ for this approach was mainly due to the memory limitations imposed by our GPU.

\subsubsection{Post-processing and submission preparations:}

For all approaches, we re-shaped the test images into $256\times256$, which the size of training data. Then, we re-shaped the predicted segments back to the original shape of CT images using re-sampling. However, we had to re-shape the predictions back to the shape of re-sampled input data before re-sizing them into the original shape. In both re-shaping methods, \texttt{INTER\_CUBIC} interpolation introduced in OpenCV~\cite{itseez2015opencv} library was used.

For all approaches in Task 1,  we used the academic version of Weights and Biases~\cite{wandb} for tracking and analyzing experiments and the corresponding performance. All the experiments with the corresponding best checkpoints are available on GitHub\footnote{\url{https://github.com/vlbthambawita/hecktor\_2022\_MLC}}.

\subsection{Task 2: Prediction of Recurrence-Free Survival}

\subsubsection{Estimation of kidney function}
We include the estimated kidney function as a feature for the XGBoost model from the third approach of Task 2 as this might improve the predictions of \gls{RFS}. 
Prior research indicates that chronic kidney disease can worsen the prognosis of cancer patients and that monitoring the kidney function of cancer patients is crucial~\cite{e2018assessment}. 

The feature is created using the Cockraft-Gault formula, which is among the most widely used formulas for estimating the kidney function~\cite{cockcroft1976KidneyFunction}. This formula requires the gender, age, body weight and serum creatinine concentration. Because serum creatinine is not available in the dataset, the average values for men and women are used instead~\cite{averageCreatinine}. Indeed, when plotting the correlation matrix for the training data, we observe that there is a positive correlation between the estimated kidney function and the \gls{RFS} (correlation = $0.26$), indicating that higher kidney function is associated with a longer time to recurrence. The entire correlation matrix for the training dataset is shown in Figure~\ref{fig:task_2_correlation}.  

\subsubsection{Description of the approaches}
For Task 2, we proposed three different approaches. The first approach used only the patient data, while the second and third approach also included features based on the image data. The image features arrived from the segmentation masks from the best approach of Task 1. Specifically, we calculated the number of pixels per class from the predicted masks in addition to the number of slices of the CT images in the z-dimension, resulting in four additional features. Moreover, the third approach used the estimated kidney function as a feature.
For all three approaches, we used $10$-fold cross-validation on the development data to determine the best hyperparameters and model.
The hyperparameters are selected based on the \gls{RMSE} of the model, which should be as low as possible.
The final models are trained on the entire training dataset using the identified set of hyperparameters.
The resulting models are then used on the test dataset to make the predictions for the challenge evaluation. All experiments are performed using the scikit-learn library~\cite{sklearn_api}.

\begin{description}
    \item{\textbf{Approach 1:}} The first approach used the Random Forest \cite{breiman2001random} algorithm to predict \gls{RFS} using only the patient data. Random Forest was chosen because it is known to work well on tabular data and is often used as a baseline for medical-related machine learning problems. All features provided in the training data were used besides the patient ID. Based on the cross-validation results (RMSE of 988.47), the hyperparameters for the Random Forest were set as the following; max features as the number of features divided by three, and the number of trees was set to 100. All other hyperparameters used the default value set by scikit-learn.

    \item{\textbf{Approach 2:}} For the second approach, we used the same algorithm as the first approach, but with additional image features as described in the beginning of the subsection. The RMSE from the cross-validation of the training data was $962.83$, which was an improvement compared to the first approach showing that the inclusion of image data has a positive effect on the results. The hyperparameters used for the Random Forest in the second approach are as follows; max features as the number of features and the number of trees 200. All other hyperparameters used the default value as set by scikit-learn. 

    \item{\textbf{Approach 3:}} Regarding the third approach, an XGBoost~\cite{Chen2016xgboost} regression model was trained to predict \gls{RFS} using the available patient data and the image features with one additional feature representing the estimated kidney function. The feature representing alcohol consumption was removed because the majority of the patients in both training and test dataset do not have any registered value for this feature. The patient ID was not included in the training dataset. The RMSE from the cross-validation on the training data was $909.09$. The hyperparameters for the XGBoost model are: `n\_estimators' = $120$, `learning\_rate' = $0.05$, `max\_depth' = $4$, `subsample' = $0.7$, `colsample\_bytree' = $0.6$, `colsample\_bynode' = $1$ and `colsample\_bylevel' = $0.8$. The other hyperparameters used the default value.    
\end{description}

\subsubsection{Investigating feature importance}
After training the XGBoost model, feature importance is explored using Shapley additive explanations (SHAP)~\cite{lundberg2019SHAP}. SHAP approximates Shapley values, which origin from game theory and assigns values to the features based on how much they contribute to the prediction~\cite{lundberg2019SHAP,Young1985ShapleyValues}. Consequently, it is possible to investigate which features the model regards as most important. 

\section{Discussion and Results}
%

\begin{figure}[!t]
    \newcommand{\imagesize}{.1135}
    \scriptsize
    \setlength{\tabcolsep}{1pt}
    
\begin{tabular}{ c c c c c c c c }
    
    Slice 1 & Slice 2 & Slice 3 & Slice 4 & Slice 5 & Slice 6 & Slice 7 & Slice 8 \\

    \raisebox{1.1\height}{\rotatebox{90}{CT}}
    \includegraphics[width=\imagesize\linewidth,height=\imagesize\linewidth]{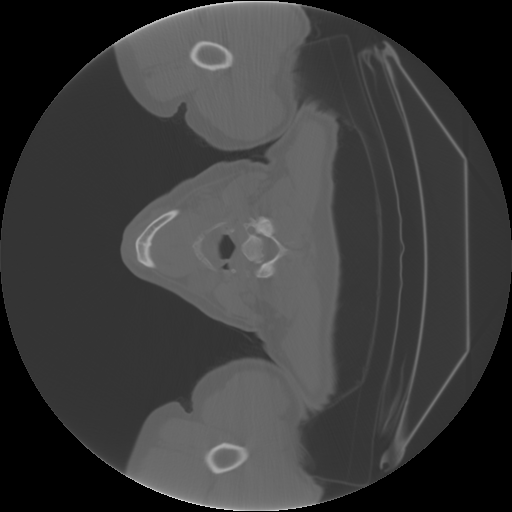} &
    \includegraphics[width=\imagesize\linewidth,height=\imagesize\linewidth]{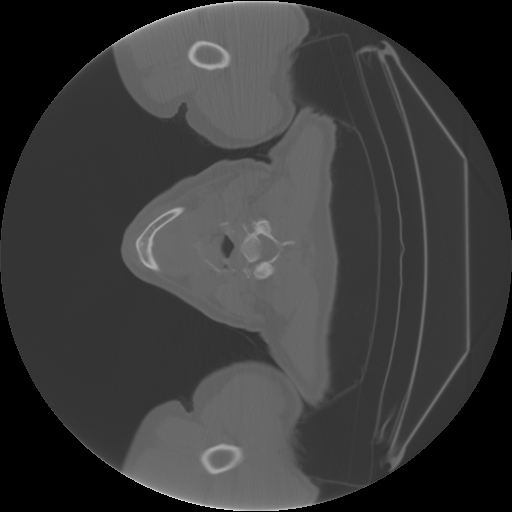} &
    \includegraphics[width=\imagesize\linewidth,height=\imagesize\linewidth]{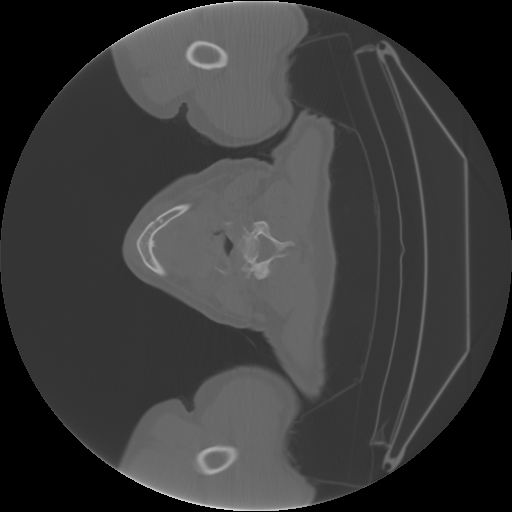} &
    \includegraphics[width=\imagesize\linewidth,height=\imagesize\linewidth]{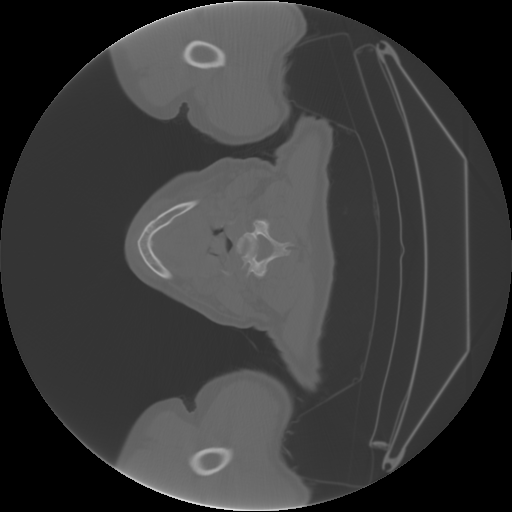} &
    \includegraphics[width=\imagesize\linewidth,height=\imagesize\linewidth]{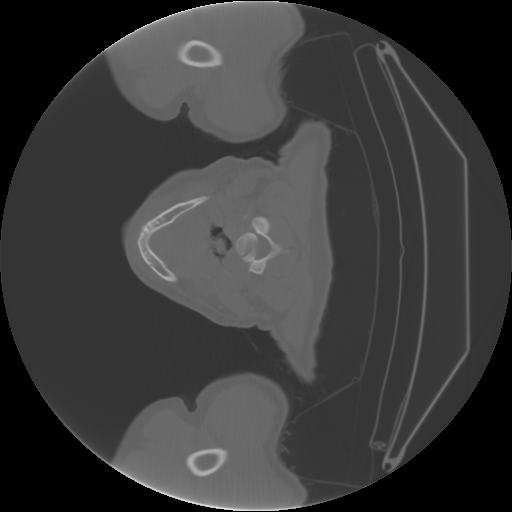} &
    \includegraphics[width=\imagesize\linewidth,height=\imagesize\linewidth]{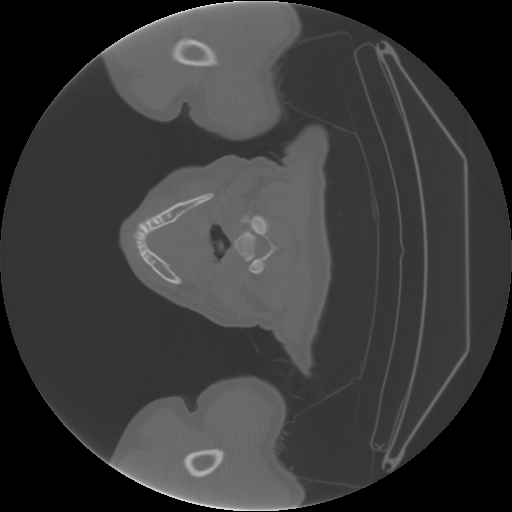} &
    \includegraphics[width=\imagesize\linewidth,height=\imagesize\linewidth]{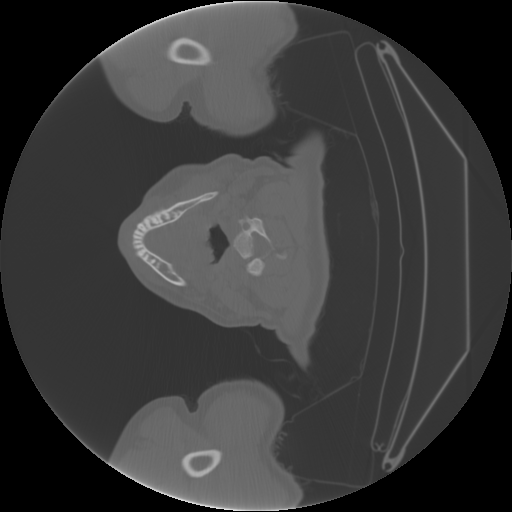} &
    \includegraphics[width=\imagesize\linewidth,height=\imagesize\linewidth]{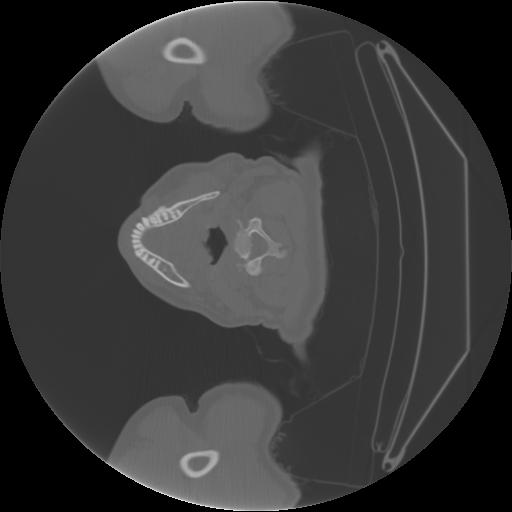} \\

    \raisebox{.6\height}{\rotatebox{90}{PET}}
    \includegraphics[width=\imagesize\linewidth,height=\imagesize\linewidth]{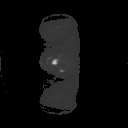} &
    \includegraphics[width=\imagesize\linewidth,height=\imagesize\linewidth]{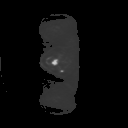} &
    \includegraphics[width=\imagesize\linewidth,height=\imagesize\linewidth]{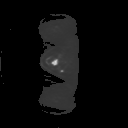} &
    \includegraphics[width=\imagesize\linewidth,height=\imagesize\linewidth]{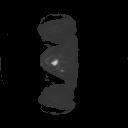} &
    \includegraphics[width=\imagesize\linewidth,height=\imagesize\linewidth]{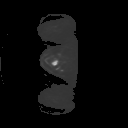} &
    \includegraphics[width=\imagesize\linewidth,height=\imagesize\linewidth]{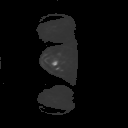} &
    \includegraphics[width=\imagesize\linewidth,height=\imagesize\linewidth]{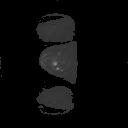} &
    \includegraphics[width=\imagesize\linewidth,height=\imagesize\linewidth]{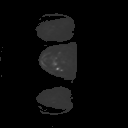} \\

    \raisebox{.2\height}{\rotatebox{90}{Model 1}}
    \includegraphics[width=\imagesize\linewidth,height=\imagesize\linewidth]{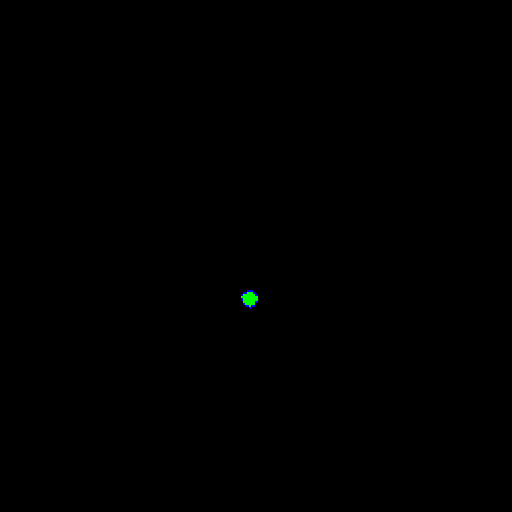} &
    \includegraphics[width=\imagesize\linewidth,height=\imagesize\linewidth]{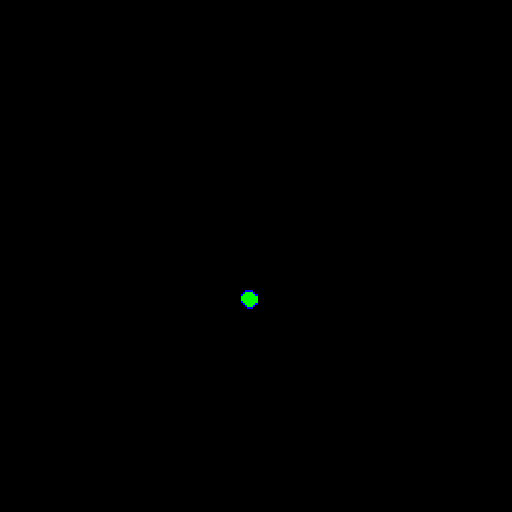} &
    \includegraphics[width=\imagesize\linewidth,height=\imagesize\linewidth]{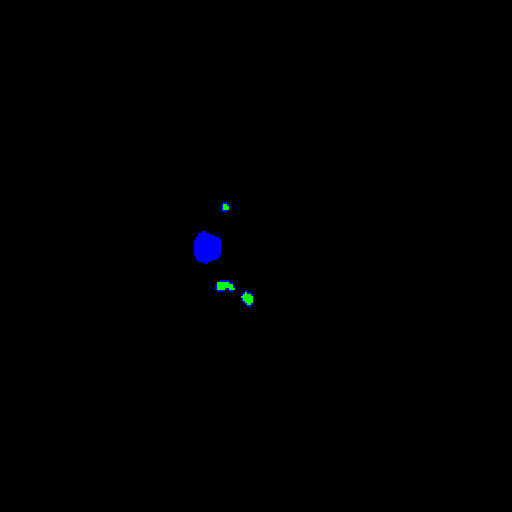} &
    \includegraphics[width=\imagesize\linewidth,height=\imagesize\linewidth]{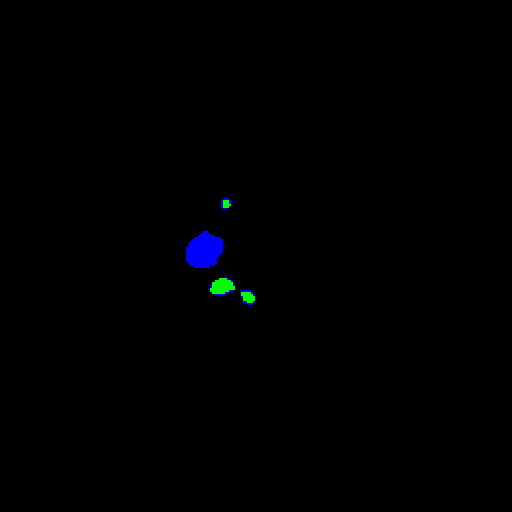} &
    \includegraphics[width=\imagesize\linewidth,height=\imagesize\linewidth]{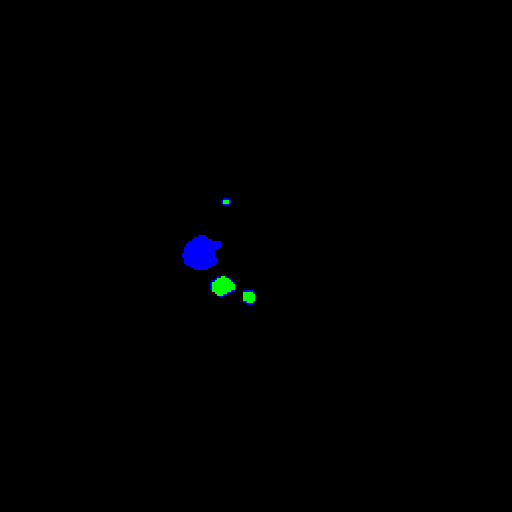} &
    \includegraphics[width=\imagesize\linewidth,height=\imagesize\linewidth]{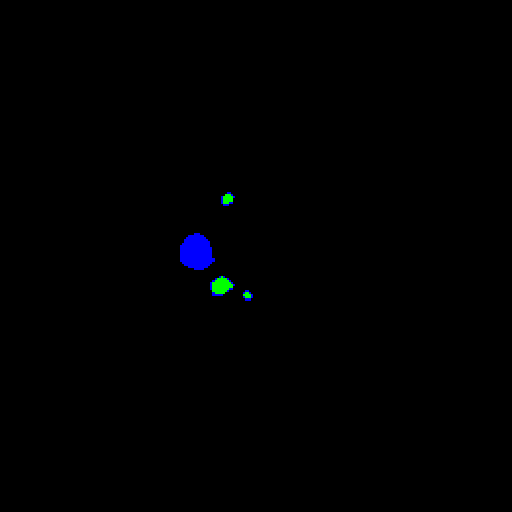} &
    \includegraphics[width=\imagesize\linewidth,height=\imagesize\linewidth]{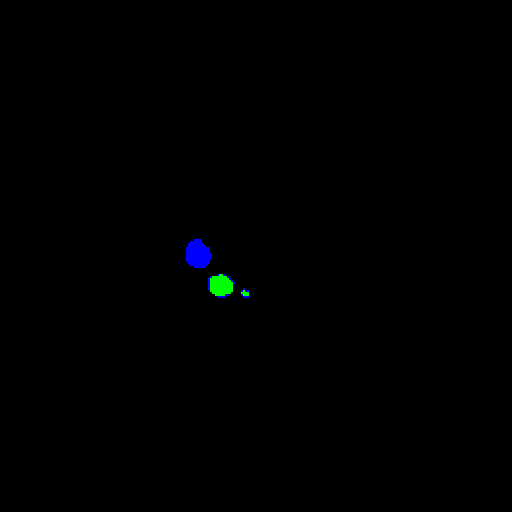} &
    \includegraphics[width=\imagesize\linewidth,height=\imagesize\linewidth]{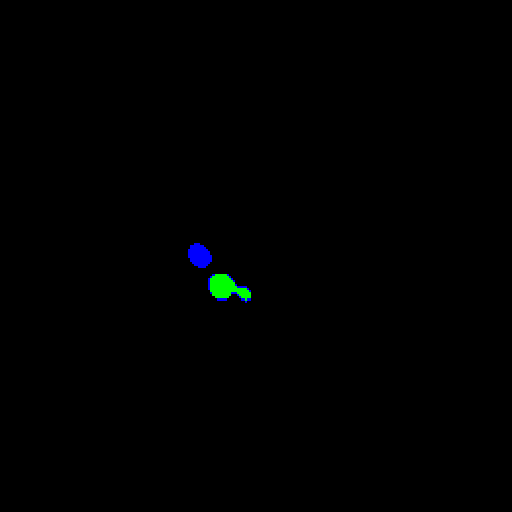} \\
    
    \raisebox{.15\height}{\rotatebox{90}{Model 2}}
    \includegraphics[width=\imagesize\linewidth,height=\imagesize\linewidth]{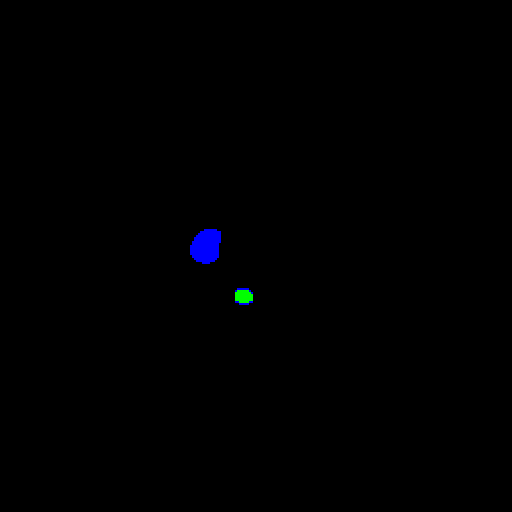} &
    \includegraphics[width=\imagesize\linewidth,height=\imagesize\linewidth]{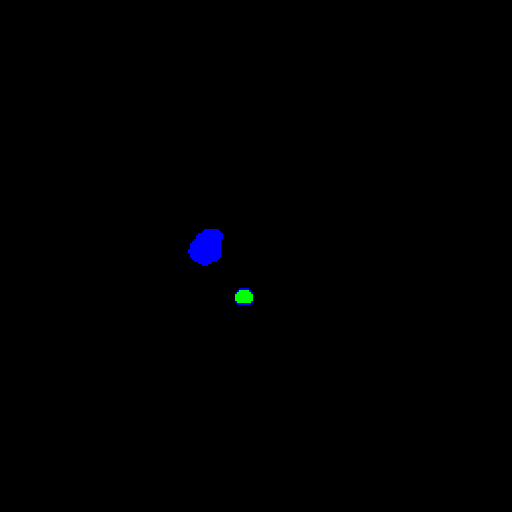} &
    \includegraphics[width=\imagesize\linewidth,height=\imagesize\linewidth]{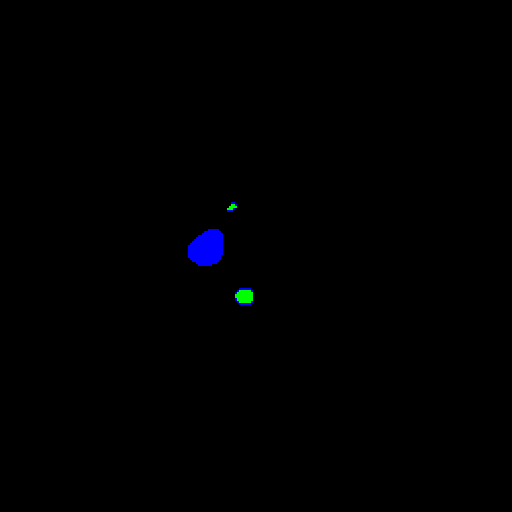} &
    \includegraphics[width=\imagesize\linewidth,height=\imagesize\linewidth]{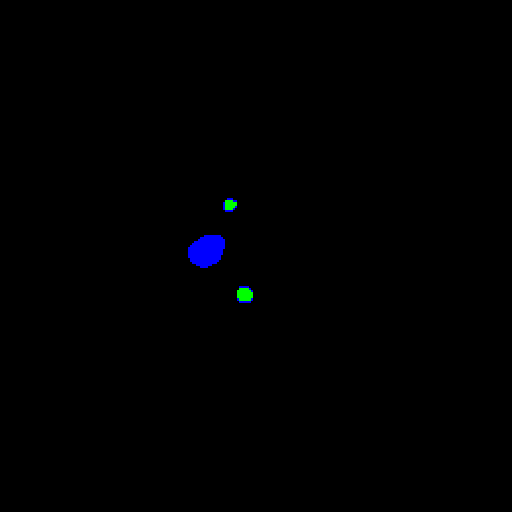} &
    \includegraphics[width=\imagesize\linewidth,height=\imagesize\linewidth]{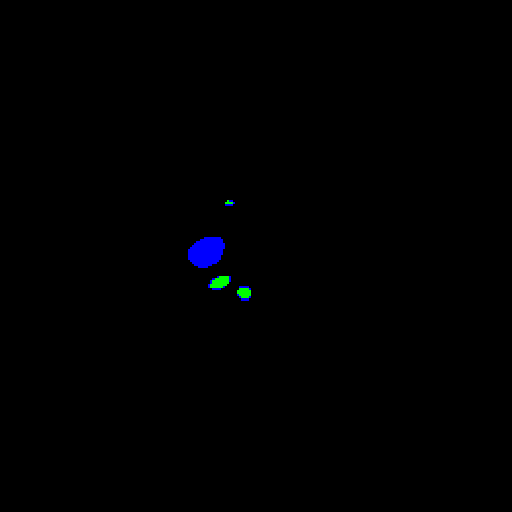} &
    \includegraphics[width=\imagesize\linewidth,height=\imagesize\linewidth]{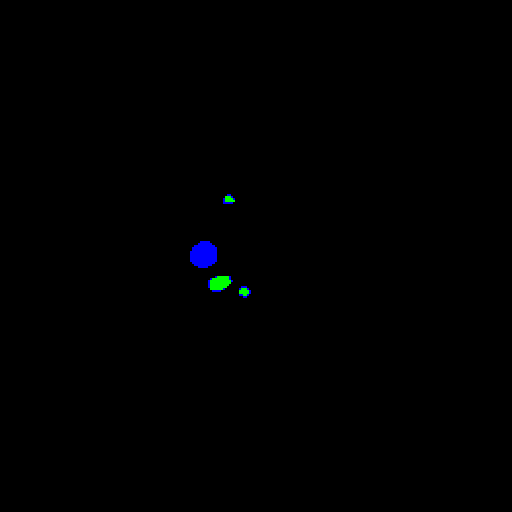} &
    \includegraphics[width=\imagesize\linewidth,height=\imagesize\linewidth]{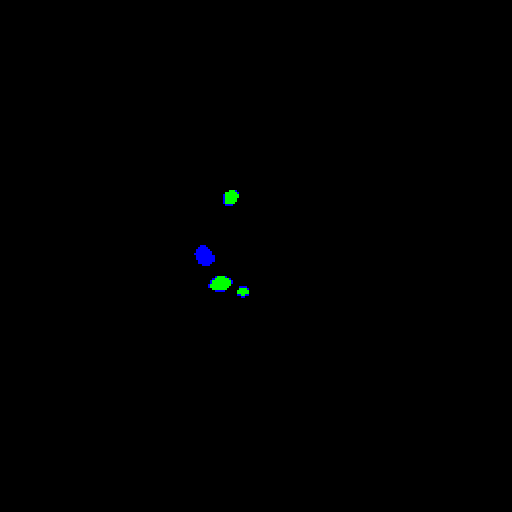} &
    \includegraphics[width=\imagesize\linewidth,height=\imagesize\linewidth]{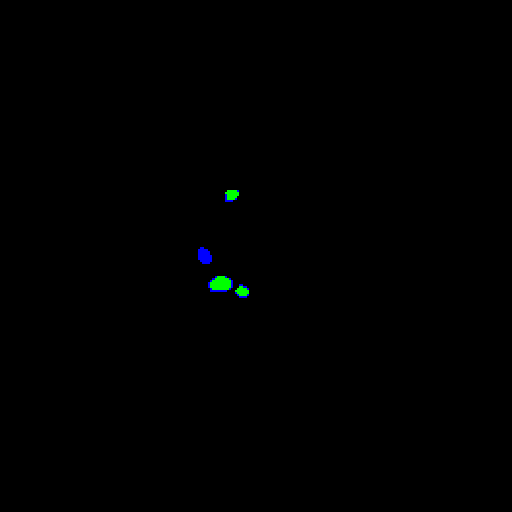} \\
    
    \raisebox{.15\height}{\rotatebox{90}{Model 3}}
    \includegraphics[width=\imagesize\linewidth,height=\imagesize\linewidth]{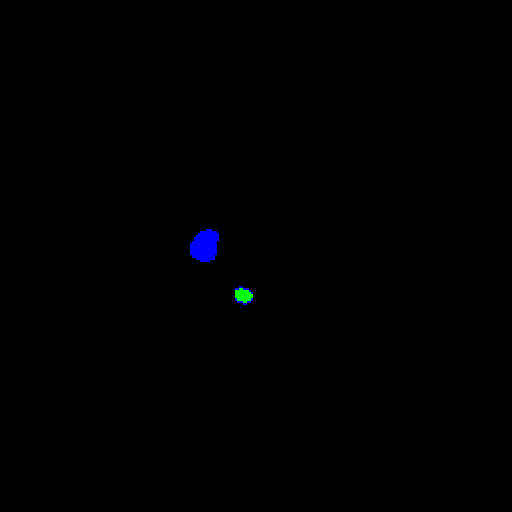} &
    \includegraphics[width=\imagesize\linewidth,height=\imagesize\linewidth]{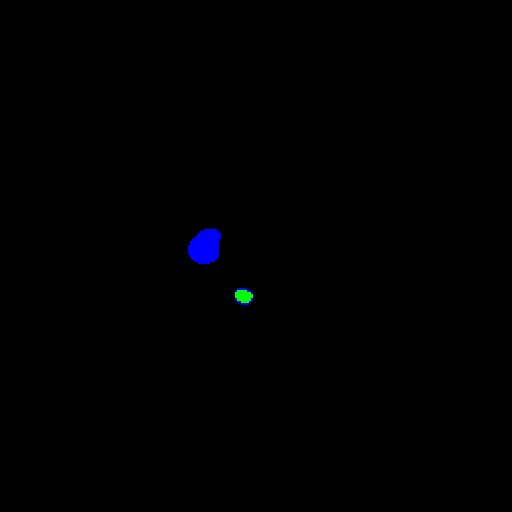} &
    \includegraphics[width=\imagesize\linewidth,height=\imagesize\linewidth]{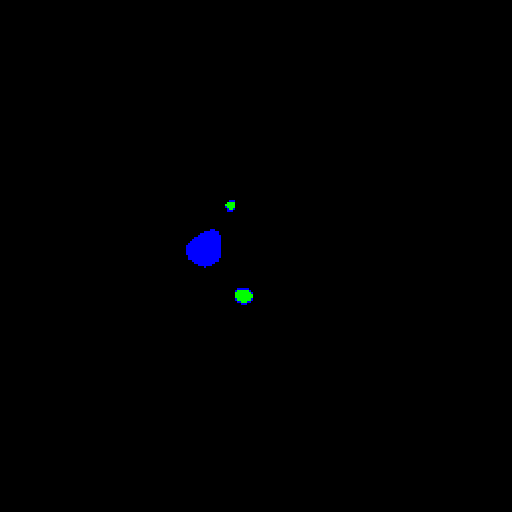} &
    \includegraphics[width=\imagesize\linewidth,height=\imagesize\linewidth]{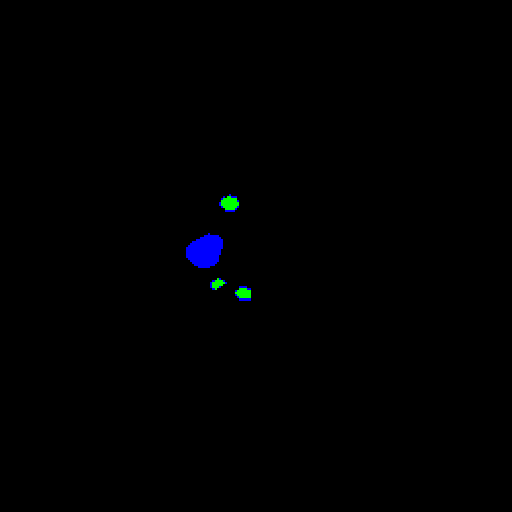} &
    \includegraphics[width=\imagesize\linewidth,height=\imagesize\linewidth]{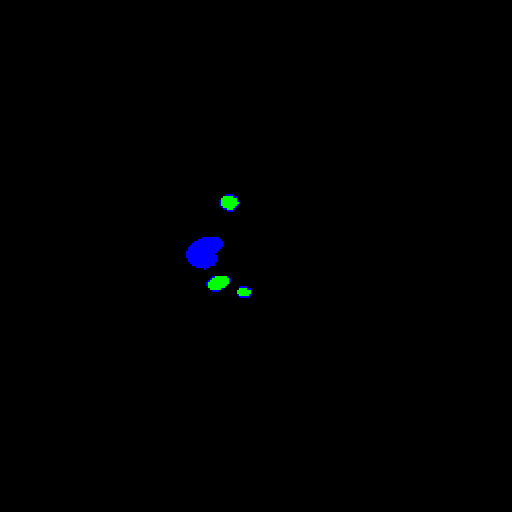} &
    \includegraphics[width=\imagesize\linewidth,height=\imagesize\linewidth]{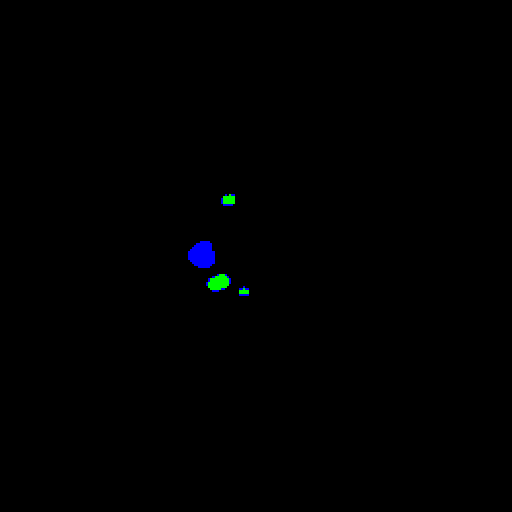} &
    \includegraphics[width=\imagesize\linewidth,height=\imagesize\linewidth]{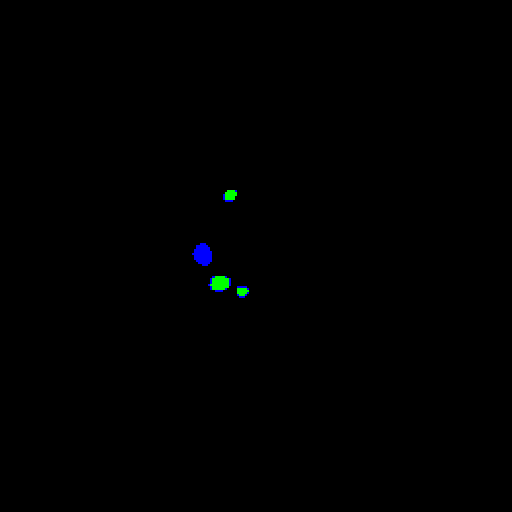} &
    \includegraphics[width=\imagesize\linewidth,height=\imagesize\linewidth]{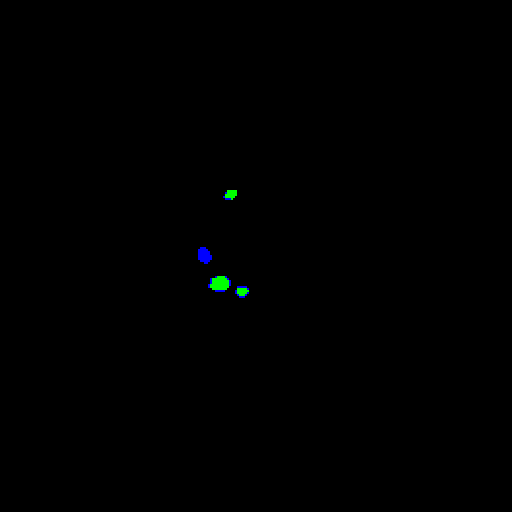} \\
    
    \raisebox{1.1\height}{\rotatebox{90}{GT}}
    \includegraphics[width=\imagesize\linewidth,height=\imagesize\linewidth]{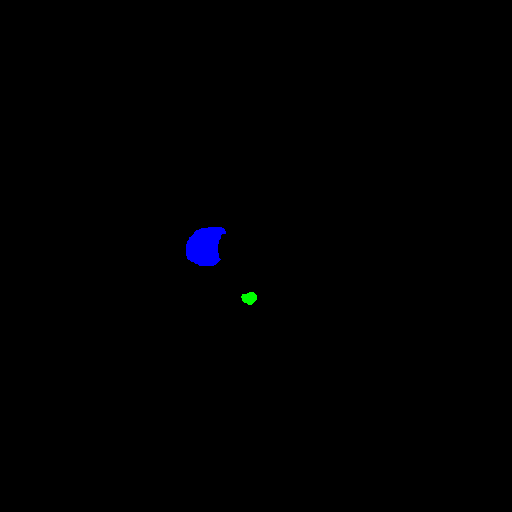} &
    \includegraphics[width=\imagesize\linewidth,height=\imagesize\linewidth]{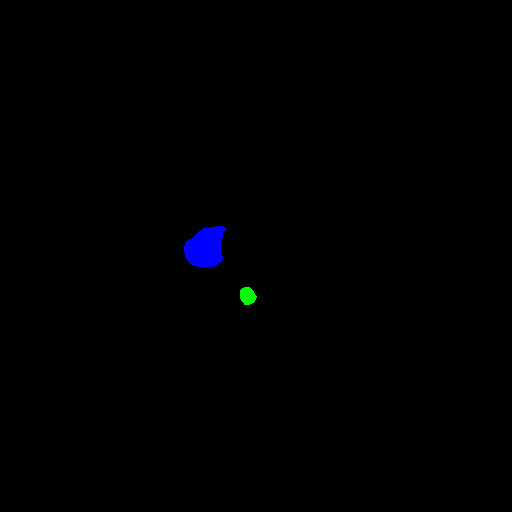} &
    \includegraphics[width=\imagesize\linewidth,height=\imagesize\linewidth]{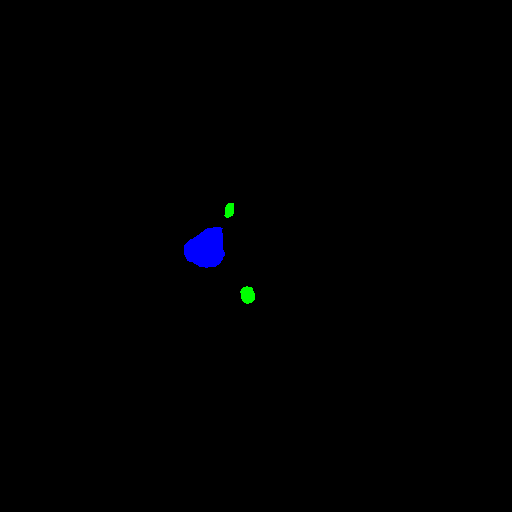} &
    \includegraphics[width=\imagesize\linewidth,height=\imagesize\linewidth]{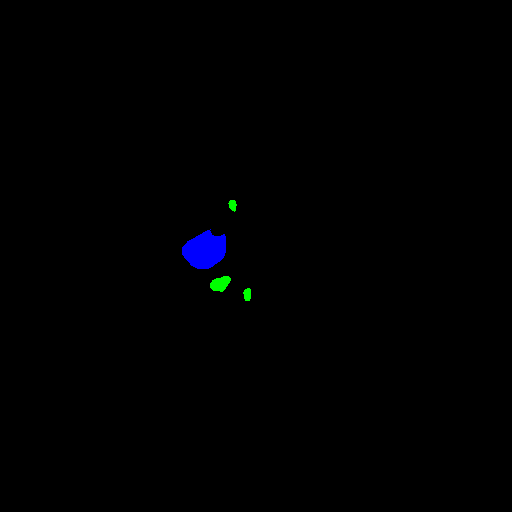} &
    \includegraphics[width=\imagesize\linewidth,height=\imagesize\linewidth]{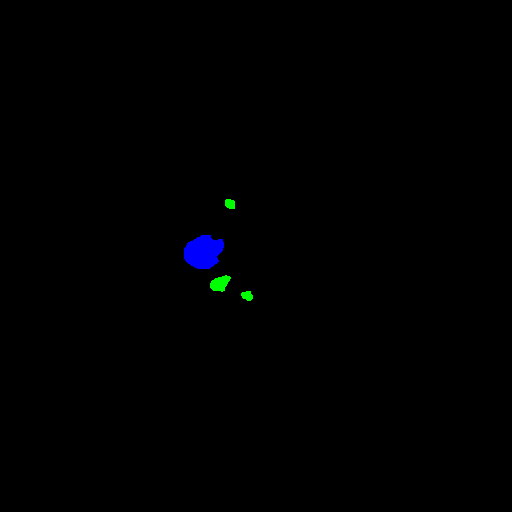} &
    \includegraphics[width=\imagesize\linewidth,height=\imagesize\linewidth]{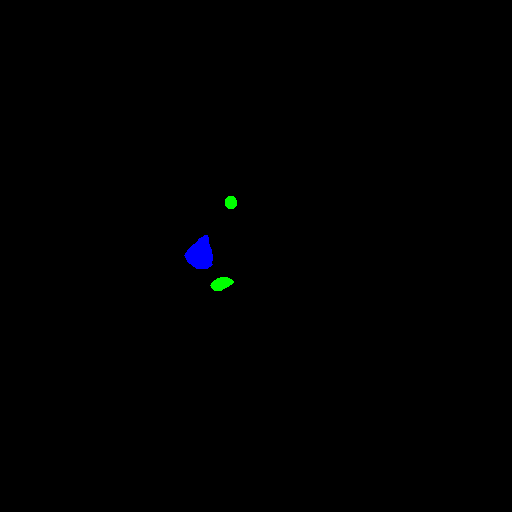} &
    \includegraphics[width=\imagesize\linewidth,height=\imagesize\linewidth]{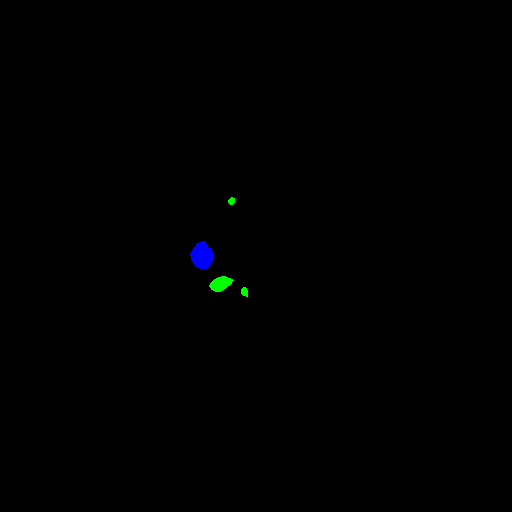} &
    \includegraphics[width=\imagesize\linewidth,height=\imagesize\linewidth]{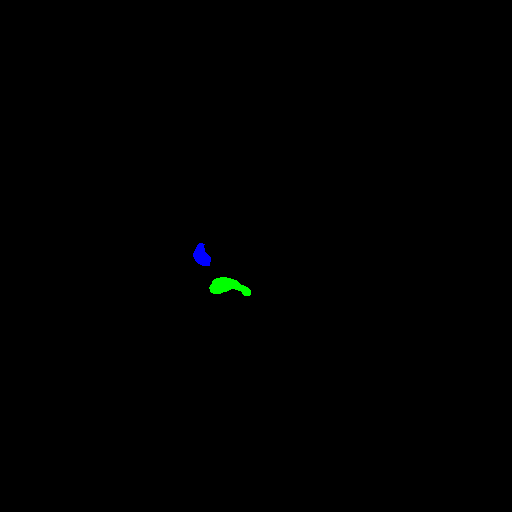} \\

    \end{tabular}
    \caption{Example predictions taken from each model at different slices and the corresponding ground truth.}\label{figure:Slice_Prediction_Examples}
\end{figure}
\begin{table}[!t]
    \centering
    \caption{Official HECKTOR Challenge 2022 results for Task 1 using the mean aggregated Dice metric for the ranking. The best values are marked using \textbf{bold text}.}\label{table:Table_Task_1_Results}
    \setlength{\tabcolsep}{5pt}
    \npdecimalsign{.}
    \nprounddigits{\decimalplaces}
    \begin{tabular}{ c c c n{1}{\decimalplaces} n{1}{\decimalplaces} n{1}{\decimalplaces} n{1}{\decimalplaces} n{1}{\decimalplaces} n{1}{\decimalplaces} }
        \toprule
        ID & Model & Dataset & \multicolumn{3}{c}{Validation} & \multicolumn{3}{c}{Test} \\
        & & & {GTVp} & {GTVn} & {Mean} & {GTVp} & {GTVn} & {Mean} \\
        \midrule
        1 & UNet++  & CT       & 0.5597 & 0.7213 & 0.6405 & 0.4659 & 0.5357 & 0.5008 \\
        2 & UNet++  & CT + PET & 0.6011 & 0.6743 & 0.6377 & 0.6065 & 0.6038 & 0.6052 \\
        3 & TriUnet & CT + PET & \textbf{0.671} & \textbf{0.722} & \textbf{0.696} & \textbf{0.659} & \textbf{0.654} & \textbf{0.657} \\
        \bottomrule
    \end{tabular}
\end{table}

\begin{table}[!t]
    \centering
    \caption{Official HECKTOR Challenge 2022 results for Task 2 using C-index for the ranking. The best values are marked using \textbf{bold text}.}\label{table:Table_Task_2_Results}
    \setlength{\tabcolsep}{10pt}
    \begin{tabular}{ c c c c }
        \toprule
        ID & Model & Data & C-index \\
        \midrule
        1 & Random Forest & Patient data & 0.585\\
        2 & Random Forest & Patient and image data & 0.589 \\
        3 & XGBoost & Patient and image data & \textbf{0.656} \\
        \bottomrule
    \end{tabular}
\end{table}

\subsection{Task 1}
Looking at the results for Task 1 in Table~\ref{table:Table_Task_1_Results}, we see that the first approach struggles to detect GTVp in both the validation and the test datasets. This is also shown in Figure~\ref{figure:Slice_Prediction_Examples}, where the first model is unable to detect the presence of GTVp until the third slice. Despite not being able to detect GTVp very well, the first approach performs well on segmenting GTVn on the validation dataset, but not on the test dataset. This can most likely be attributed to the differences between the validation and test datasets, as the other approaches show similar results. Adding information from the PET scans for the latter two approaches seems to help in detecting GTVp, as evident by the improved scores in Table~\ref{table:Table_Task_1_Results}, and they are both able to detect GTVp in all eight slices from the example in Figure~\ref{figure:Slice_Prediction_Examples}. The differences between Approaches 2 and 3 indicate that extracting features from the CT and PET scans independently seems to be the most suitable technique. 


\subsection{Task 2}

\begin{figure}[!t]
    \centering
    \begin{tikzpicture}
      \begin{axis}[
        xbar,
        bar width = .2cm,
        y axis line style = { opacity = 0 },
        axis x line       = none,
        tickwidth         = 0pt,
        ytick             = data,
        enlarge y limits  = 0.08,
        enlarge x limits  = 0.02,
        symbolic y coords = {
            Tobacco, CenterID, HPVstatus, eGFR, Weight, 
            Performance status, Count 1, Count 0, Count 2, Age,
            Surgery, dim z, Gender, Chemotherapy },
        nodes near coords,
      ]
      \addplot coordinates { (221.44229552224851,Tobacco) (157.16983430658996,CenterID)
                             (149.41783634574102,HPVstatus)  (89.15552515411888,eGFR)
                             (77.93642775772419,Weight)  (77.31576789744791,Performance status)
                             (69.86924467891895,Count 1)  (66.72384837570927,Count 0)
                             (55.872933522133174,Count 2)  (47.550111878156045,Age)
                             (42.24799618962943,Surgery)  (33.36418964711438,dim z)
                             (27.228049502727902,Gender)  (11.07081818789859,Chemotherapy)};
      \legend{SHAP Value}
      \end{axis}
    \end{tikzpicture}
    \caption{Feature importance for the XGBoost model from Task 2, trained on image features and patient information. Count 0, Count 1, Count 2 and dim z are image features.}
    \label{fig:task_2_SHAP}
\end{figure}
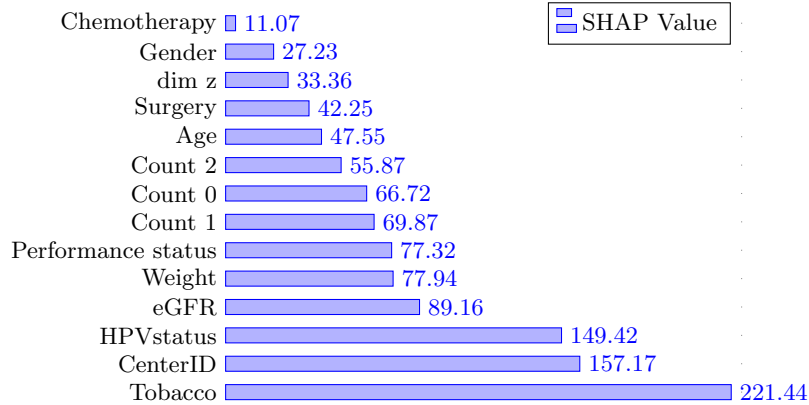



 
 

For Task 2, the results can be seen in Table \ref{table:Table_Task_2_Results}. From the results, we can observe several interesting insights. Firstly, adding additional data to the patient data gives better predictions. This can be observed in the difference between approach 1 and 2 when image data was added as additional features. We can also observe that XGBoost outperforms Random Forest by a large margin. This correlates with general findings in the literature that XGBoost is one of the best working methods. This questions also the general concept of using Random Forest as a baseline and suggest that in general it should be replaced with XGBoost instead.

\begin{figure}[!t]
    \centering
    \begin{tikzpicture}
    \begin{axis}[
        width=\textwidth,
        height=.7\textwidth,
        axis equal image, 
        scatter, 
        colormap/viridis, 
        colorbar, 
        point meta min=-1,
        point meta max=1,
        axis line style={draw=none},
        ymin=0, ymax=14,
        xmin=0, xmax=14,
        yticklabels={Count 0, Count 1, Count 2, dim z,
                CenterID, Gender, Age, Weight, Tobacco,
                Performance status, HPVstatus, Surgery,
                Chemotherapy, eGFR, RFS},
        ytick={0,...,15},
        yticklabel style = {xshift=-0.25cm},
        tickwidth=0pt, 
        y dir=reverse, 
        xticklabel pos=bottom, 
        xticklabel style = {yshift=-0.25cm,rotate=45, anchor=north east, inner sep=0mm},
        xticklabels={Count 0, Count 1, Count 2, dim z,
                CenterID, Gender, Age, Weight, Tobacco,
                Performance status, HPVstatus, Surgery,
                Chemotherapy, eGFR, RFS},
        xtick={0,...,15},
        enlargelimits={abs=0.7pt}, 
    ]
    \addplot +[
        point meta=explicit, 
        mark=square*,
        mark size=6pt,
        only marks, 
    ] table [ meta=value ]{data/Correlation.dat};
    \end{axis}
    \end{tikzpicture}
        \caption{Correlation matrix for the training data applied in Task 2 for the third approach. Count 0, Count 1, Count 2, and dim z are image features.}
        \label{fig:task_2_correlation}
\end{figure}
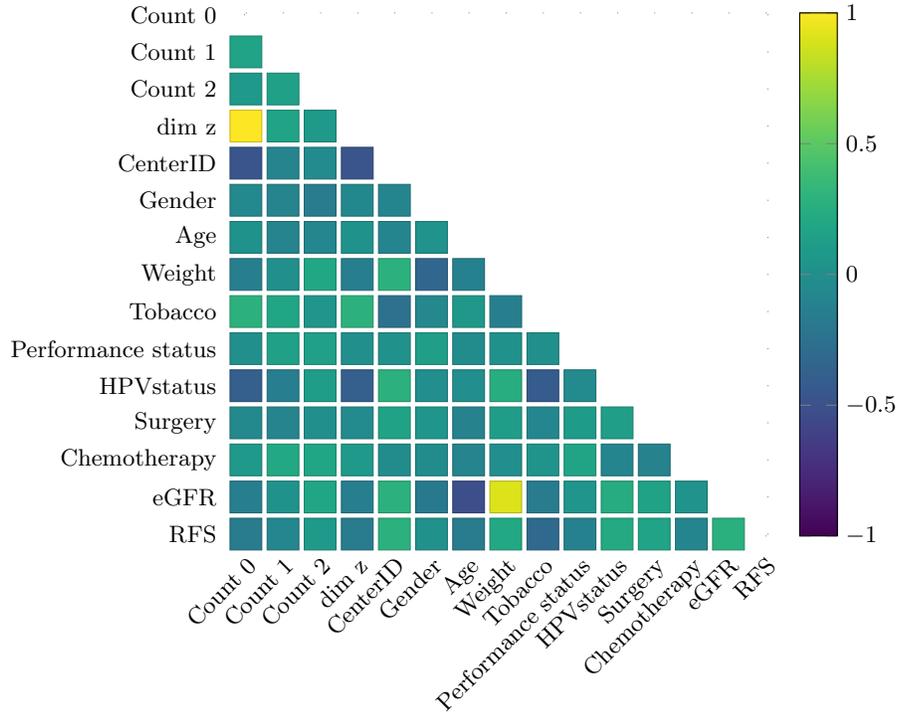


The SHAP values estimated for the third approach are plotted in Figure~\ref{fig:task_2_SHAP} and give us a better understanding of which features are most relevant to the model. We observe that the top five features are Tobacco, CenterID, HPVstatus, estimated glomerular filtration rate (eGFR) which represents the kidney function, and Weight (most important first). Tobacco consumption is the most important feature. This is not surprising as it is well-known that tobacco increases the risk of developing head and neck cancer, see for example~\cite{Hashibe2009TobaccoHeadNeckCancer}. The kidney function is ranked as number four and seems to be an important indicator for RFS. This finding is in line with earlier research, where a relationship between kidney function and prognoses for cancer patients has been identified~\cite{e2018assessment}. An important limitation is that the true serum creatinine values were not available in the provided dataset. The creatinine concentration will to a large extent affect the estimated kidney function, and only applying the average gender values is not enough for getting accurate individual estimations. Despite this, we believe that serum creatinine should be included in future datasets to see if the model performance can be further improved. Interestingly, CenterID is ranked as the second most important feature, which is also confirmed by a positive correlation between the CenterID and RFS in the correlation plot (Figure~\ref{fig:task_2_correlation}). These observations might be due to different patient populations at different centers, e.g., there might be more severely ill patients treated at one center while less serious cases are treated at another center. Another possibility is that the medical doctors choose different strategies for treating the patients or that the surgical skills differ between the centers. However, neither Surgery or Chemotherapy are among the highest-ranked features. The CenterID was not encoded as a categorical feature. If this had been done, the results from the SHAP analysis might change. The five least important features are Chemotherapy, Gender, dim z, surgery, and Age (least important first). The image features rank in the middle range regarding feature importance showing that they can be an important factor in predicting RFS. Considering that our imaging method and the image features are simple, we assume that with more advanced image analysis methods and the corresponding resulting features, the importance of the image-related features might increase even further.


\section{Conclusion}
In conclusion, our simple methods were not able to perform at the same levels as the highest rankest submissions despite achieving reasonable results. We got some interesting insights regarding combinations of different data modalities showing that the combination of different sources improves the results even when simple methods are used. For Task 2, we also had a closer look at feature importance, revealing some interesting features such as the usefulness of eGFR. For future work, it would be interesting to apply the feature importance analysis to other solutions of the competition to investigate if they are leading to similar findings. Furthermore, we would also like to investigate the CenterID correlation to explore if a hospital-specific treatment or country-related factor is influencing it.

\bibliographystyle{splncs04}
\bibliography{bibliography}

\end{document}